\documentclass[preprint,aps,showpacs]{revtex4}
\def\beq{\begin{equation}}
\def\eeq{\end{equation}}
\def\bea{\begin{eqnarray}}
\def\eea{\end{eqnarray}}
\def\bwi{\begin{widetext}}
\def\ewi{\end{widetext}}

\def\by{\over}

\def\tst{\textstyle}

\def\eno#1{Eq.~(\ref{#1})}
\def\Eno#1{Equation (\ref{#1})}
\def\etwo#1#2{Eqs.~(\ref{#1}) and (\ref{#2})}

\def\fnm{\footnotemark}
\def\fnt{\footnotetext}

\def\gam{\gamma}
\def\dta{\delta}
\def\eps{\epsilon}
\def\tta{\theta}

\def\lam{\lambda}

\def\om{\omega}

\def\Om{\Omega}

\def\Tr{{\rm Tr}}
\def\grad{\nabla}
\def\apx{\approx}
\def\ptl{\partial}

\def\hf{{1\over2}}
\def\tshf{\tst\hf}

\def\tofro{\leftrightarrow}

\def\mixpar#1#2#3{{\ptl^2#1 \by \ptl#2\ptl#3}}

\def\lp{\left(}
\def\rp{\right)}
\def\lsb{\left[}
\def\rsb{\right]}

\def\ham{{\cal H}}
\def\ket#1{|#1\rangle}

\def\tran#1#2{\langle#1|#2\rangle}

\def\mel#1#2#3{\langle#1|#2|#3\rangle}

\def\ba{{\bf a}}
\def\bJ{{\bf J}}

\def\xhat{{\bf{\hat x}}}
\def\yhat{{\bf{\hat y}}}
\def\zhat{{\bf{\hat z}}}
\def\nhat{{\bf{\hat n}}}

\def\Jxop{J_x^{\rm op}}
\def\Jyop{J_y^{\rm op}}
\def\Jzop{J_z^{\rm op}}
\def\Jiop{J_i^{\rm op}}
\def\bJop{\bJ^{\rm op}}

\def\Jmi{J^{\rm op}_-}
\def\ylm{Y_{\ell m}}
\def\ylmop{{\cal Y}_{\ell m}}
\def\jtil{{\tilde\jmath}}

\def\zbar{{\bar z}}
\def\baz{{\bar z}}

\def\bzi{{\bar z}_i}
\def\bzf{{\bar z}_f}
\def\zbf{{\bar z}_f}

\def\Ksc{K_{{\rm sc}}}
\def\Hsc{H_{{\rm sc}}}
\def\Fsc{F_{{\rm sc}}}

\def\Gsc{G_{{\rm sc}}}

\def\Zsc{Z_{{\rm sc}}}
\def\Asc{{\cal A}_{\rm sc}^+}
\def\Ac{{\cal A}_c^+}
\def\zsc{z_{{\rm sc}}}
\def\bzsc{\zbar_{{\rm sc}}}

\def\Fe8{Fe$_8$}

\begin{document}


\title{Bohr-Sommerfeld Quantization of Spin Hamiltonians}

\author{Anupam Garg}
\email[e-mail address: ]{agarg@northwestern.edu}
\affiliation{Department of Physics and Astronomy, Northwestern University,
Evanston, Illinois 60208}

\author{Michael Stone}
\affiliation{Department of Physics, University of Illinois
at Urbana-Champaign, Urbana, IL 61801}

\date{\today}

\begin{abstract}
The Bohr-Sommerfeld rule for a spin system is obtained, including the
first quantum corrections. The rule applies to both integer and half-integer
spin, and respects Kramers degeneracy for time-reversal invariant systems.
It is tested for various models, in particular the Lipkin-Meshkov-Glick
model, and found to agree very well with exact results.
\end{abstract}


\maketitle

The question in this Letter was investigated twentysome years ago by
R.~Shankar, in a paper in this journal with a nearly identical title
\cite{shankar80}. Many physical systems involve large-magnitude
spin or pseudospin degrees of freedom. Examples include spinning
molecules \cite{harter77+84,robbins89}, superdeformed rotating
nuclei \cite{hamamoto94+95}, magnetic molecular solids
\cite{villain00}, and the Lipkin-Meshkov-Glick (LMG) model
of certain collective excitations in nuclei \cite{lipkin65}. In all these
cases, because the spin magnitude, $j$, is large, a semiclassical
approach is natural. In some cases, such an approach is effective
even for $j$ as small as $1/2$, e.g., in studies of magnetic
order \cite{mattis, haldane83+85, henley98}.

It may therefore surprise some readers that
so far we have lacked even the Bohr-Sommerfeld (BS) quantization rule
for a single spin. All previous authors cited have used heuristic or
ad hoc BS rules that vary in their treatment of the first quantum
corrections \cite{qtmcorrect}. In this paper, we use the semiclassical
propagator for spin, which has only recently been fully understood
\cite{spg+etal}, to find the BS rule systematically. We shall first
state the rule, then apply it to a few model problems, and finally prove it.

The general problem is as follows. We are given a quantum Hamiltonian 
\beq
\ham = \ham(\Jxop, \Jyop, \Jzop),
\eeq
where $\Jiop$ ($i = x$, $y$, $z$) are components of the dimensionless
spin operator, $\bJop$, with the commutator
$[\Jiop,J_j^{\rm op}] = i\eps_{ijk}J_k^{\rm op}$,
and $\ham$ is a polynomial in $\Jiop$ \cite{degreepoly}. Since
$\bJop\cdot\bJop = j(j+1)$ is a constant of
motion, the spin magnitude $j$ must also be given. We want the eigenvalues
of $\ham$ when $j$ is large, and the problem is, intuitively,
quasiclassical.

The first step is to understand the classical limit. We expect $\ham$
to correspond to some classical Hamiltonian $H_c(\bJ)$ of a c-number
angular momentum $\bJ$, with dynamics defined by the Poisson brackets
$\{J_i, J_j\} = \eps_{ijk} J_k$. Since $\bJ\cdot\bJ$ is conserved,
we can regard the motion as taking place on a sphere of radius
$j = |\bJ|$. All orbits on this sphere are closed and non intersecting,
so it may regarded as the phase space of the system. With polar
coordinates $\tta$ and $\phi$, and the usual identification of $J_z$
and $\phi$ as canonically conjugate momentum and coordinate, we can
expect the BS rule to quantize the action integral $\oint \cos\tta d\phi$.
Robbins et al{.} \cite{robbins89} set this integral to
$(2n + 1)\pi/(j+\hf)$, Harter and Patterson \cite{harter77+84} set
it to $2n \pi/[j(j+1)]^{1/2}$, and Shankar \cite{shankar80} equates
$j\Asc$ [see paragraph below \eno{bsrule2}] to $2n \pi$. Neither rule is
formulated in a context that extends to half-integer $j$.
The rule we find is
\beq
(j+{\tshf}) {\cal A}_c^+(E) + {1\by 2j} \int_0^{T(E)} \bJ\cdot\grad_J H_c dt
     = (2n+1)\pi.
      \label{bsrule}
\eeq
Here, $n$ is an integer ranging from 0 to $2j$, and
\beq
{\cal A}_c^+(E) = \oint_{H_c = E} (1 - \cos\tta) d\phi
    \quad ({\rm mod}\  4\pi) \label{area}
\eeq
is the area enclosed by an orbit of energy $E$ \cite{areasign}. For future
use, let us denote the two terms in \eno{bsrule} by $I_1$ and $I_2$, and
also define
\beq
\jtil = j + \tshf.
\eeq
The term $I_2$ is also evaluated along the orbit $H_c =E$, which is
taken to have a period $T(E)$. This term, and the extra $\hf$ in the
$j+\hf$ factor in $I_1$ represent the first quantum corrections.

To make \eno{bsrule} precise, we must also state the rule for
associating $H_c$ with $\ham$. We do this by writing
\beq
\ham(\bJop) = \sum_{\ell = 0}^{2j} \sum_{m=-j}^j
                   c_{\ell m} \ylmop(\bJop),
\eeq
where $\ylmop$ are the spherical harmonic tensor operators, and the
coefficients $c_{lm}$ are uniquely determined since
$\Tr(\ylmop^{\dagger}{\cal Y}_{\ell' m'})
          \propto \dta_{\ell\ell'}\dta_{mm'}.$
Then,
\beq
H_c(\bJ) = \sum_{\ell m} c_{\ell m} \ylm(\bJ)
\eeq
with the same $c_{\ell m}$'s, and the $\ylm$'s are solid harmonics.
Along with the $\ylmop$'s, they may be defined by the generating function
\cite{schwinger}
\beq
\lsb\begin{array}{c}
    (\ba\cdot\bJ)^{\ell} \\ (\ba\cdot\bJop)^{\ell} \end{array} \rsb
  =
%
%
   \sqrt{4\pi\by 2\ell + 1}\sum_{m=-\ell}^{\ell} 
              a_{\ell m} \lam^m
     \lsb\begin{array}{c}
          \ylm(\bJ) \\ \ylmop(\bJop) \end{array} \rsb,
    \label{ylmgenfn}
\eeq
where, $a_{\ell m} = \ell!/ [(\ell+m)! (\ell -m)!]^{1/2}$, and
\beq
\ba = \zhat - {\lam\by 2}(\xhat + i \yhat)
             + {1\by 2\lam}(\xhat - i \yhat)
\eeq
is a complex vector with $\ba\cdot\ba = 0$. Note that
given $\ham$, $H_c$ is completely determined. To get $H_c$ from $\ham$, we
also need to give the value of $j$. Hence, the $\ham \tofro H_c$
correspondence is one-to-one. Note also that the trinomial expansion of
$(\ba\cdot\bJop)^{\ell}$ is analogous to the expansion of $(aq+bp)^n$
that gives Weyl ordering.

Let us now apply \eno{bsrule} to a few models. Our first example is
$\ham = \om\Jzop$, and $H_c = \om J_z$. The
orbits are given by $\cos\tta = E/\om j = {\rm const}$, and $\dot\phi = \om$.
Hence, $I_1 = 2\jtil\pi(1-E/j\om)$, and $I_2 = \pi E/j\om$, so the
quantization condition is $E_n =\om(j-n)\pi$, with $n = 0$, 1, 2, $\ldots$,
$2j$. In this simple case, the rule is exact.

Our next example is somewhat less trivial, $\ham = \nu(\Jzop)^2$,
$H_c = \nu J_z^2$. Again $\cos\tta = {\rm const}$, $E = \nu j^2 \cos^2\tta$,
and $\dot\phi = 2\nu j \cos\tta$. The rule again yields $j\cos\tta = (j -n)$,
i.e., $E_n = \nu (j -n)^2$. This is also exact, but more importantly, it is
in accord with Kramers' theorem. It is not hard to see that this is
true for any time-reversal invariant $\ham$, for which $H_c(-\bJ) =H_c(\bJ)$.

As our third example, we consider the LMG model, just as done by Shankar.
The Hamiltonian now is
\beq
\ham = \Jzop + {r\by 2j}\lsb (\Jxop)^2 - (\Jyop)^2 \rsb,
     \label{h_lmg}
\eeq
and $H_c$ is obtained by simply deleting the `op' suffixes.
On the orbit with energy $E$,
\beq
\cos\tta = {1 \pm [1 + r\cos 2\phi(r\cos 2\phi - 2E/j)]^{1/2}
                    \by r\cos 2\phi}. \label{ttavsphi}
\eeq
Now, $H_c \to -H_c$ under a 180$^{\circ}$ rotation about $\xhat \pm \yhat$,
i.e., $\tta \to \pi - \tta$, $\phi \to \phi + \pi/2$. 
This symmetry forces states to occur in pairs, $E$ and $-E$, plus 
a nondegenerate state at $E = 0$ for integer $j$. For orbits related by
this symmetry, we have $I_1 \to 4\pi\jtil - I_1$, and $I_2 \to -I_2$,
so the BS spectrum is also symmetric about $E=0$. For
the orbit at $E=0$, $I_1 = (2j+1)\pi$, and $I_2 = 0$, so $E=0$ is an
allowed energy for integer $j$ only. From now on, we consider only
$E \ge 0$.

The energy landscape for this model looks like this. If $r \le 1$, $H_c$
has a maximum at $\tta=0$ ($H_c =j$) and a minimum at $\tta = \pi$. If
$r > 1$, these points turn into saddle points, and $H_c$ develops two
degenerate maxima [with $H_c = \hf j(r + r^{-1}$)] along the $\phi = 0$ and
$\phi=\pi$ meridians. Orbits at slightly lower energies circle these
maxima, and are separated by a separatrix at $E = j$ passing through the
north pole. (See Fig. 1 in Ref. \onlinecite{shankar80}.) Hence for $r>1$,
we expect to have pairs of levels split by tunneling for $E$ above or
around $j$, and single levels below. In this paper, due to limited space,
we will not incorportate tunneling effects into our BS calculations.
In principle this may be done by allowing the orbits to become complex
\cite{robbins89}. We have discussed how to find ground pair splittings
from the propagator elsewhere \cite{gkps}. For the orbits with $E > j$,
both signs in \eno{ttavsphi} are valid, but for $E < j$ (and any $r$)
only the minus sign is meaningful. 

\begin{table}[t]
\caption{\label{tab1}Positive part of energy spectrum of the LMG model
from the BS rule (\ref{bsrule}) and numerical diagonalization for $j=15$.}
\begin{ruledtabular}
\begin{tabular}{c c c c c c}
\multicolumn{2}{c} {$r = 0.6$} &
     \multicolumn{2}{c} {$r = 1.0$} &
           \multicolumn{2}{c}  {$r = 5.0$} \\
\mbox{BS} & \mbox{exact} & \mbox{BS} & \mbox{exact}
                           & \mbox{BS} & \mbox{exact} \\
\hline
   15.10\fnm[1] & 15.09  & 15.33\fnm[2] & 15.31  & 37.98 & 38.05 \\
   14.26 & 14.26  & 14.77\fnm[2]  & 14.80  & 37.98 & 38.05 \\
   13.38 & 13.38  & 14.12 & 14.09  & 31.37 & 31.44 \\
   12.46 & 12.46  & 13.28 & 13.27  & 31.37 & 31.44 \\
   11.52 & 11.51  & 12.38 & 12.37  & 25.35 & 25.42 \\
   10.54 & 10.54  & 11.41 & 11.41  & 25.35 & 25.42 \\
    9.55 &  9.54  & 10.40 & 10.39  & 20.02 & 20.14 \\
    8.53 &  8.53  &  9.34 &  9.33  & 20.02 & 20.05 \\
    7.50 &  7.50  &  8.24 &  8.24  & 15.66 & 16.13 \\
    6.45 &  6.45  &  7.12 &  7.11  & 15.01\fnm[1] & 15.24 \\
    5.39 &  5.39  &  5.97 &  5.96  & 12.82 & 12.63 \\
    4.33 &  4.33  &  4.80 &  4.79  & 10.46 & 10.47 \\
    3.25 &  3.25  &  3.61 &  3.61  &  7.96 &  7.93 \\
    2.17 &  2.17  &  2.41 &  2.41  &  5.36 &  5.35 \\
    1.09 &  1.09  &  1.21 &  1.21  &  2.69 &  2.69 \\
    0.00 &  0.00  &  0.00 &  0.00  &  0.00 &  0.00 \\
\end{tabular}
\end{ruledtabular}
\fnt[1]{By extrapolation}
\fnt[2]{By mapping to $q$, $p$ variables}
\end{table}

The evaluation of the two action terms in \eno{bsrule} requires simple
one-dimensional numerical integration. The resulting BS spectrum is
compared with the exact one (from numerical diagonalization of $\ham$)
in Table \ref{tab1}. The values of $j$ and $r$ are the same as those used
by Shankar, and it is evident that our BS analysis improves on his. Indeed,
it is rather good almost uniformly,
but a few aspects call for comment. For $r<1$, the highest energy exceeds
$j$, and can be found by extrapolating the BS action. This amounts to
allowing for complex orbits. For $r=1$, the energy is not quadratic
in deviations about the maximum at $\tta= 0$. In terms of the stationary
phase approximation on which \eno{bsrule} is premised (see below), this
corresponds to an exceptional case where we must include fluctuations
higher than second order (Gaussian) about the stationary phase point.
The two highest energies in Table \ref{tab1} were obtained via the mapping
$\Jxop \apx q$, $\Jyop \apx jp$, with $[q,p] =i$, and textbook BS
quantization for a particle in one dimension. The same limitation on
\eno{bsrule} is present for $r>1$  close to the tunneling barrier, i.e.,
$E \simeq j$. Now, one may have a pair of tunnel split levels such that
one level is below the barrier and the other above. Literal use of
\eno{bsrule} then yields two degenerate levels just below the barrier,
with large errors. We have chosen to find the unbound partner by
extrapolating the action for $E < j$. A proper semiclassical approach
would require including complex orbits, and non-Gaussian fluctuations.
These aspects are not unique to spin, and also occur with one-dimensional
double well potentials.

In the rest of the paper, we show how we derive the BS rule (\ref{bsrule}).
Let $\ket{\nhat}$ be a spin coherent state with maximal spin projection
along the direction $\nhat$ with polar coordinates $(\tta,\phi)$, i.e.,
$\bJop\cdot\nhat\ket{\nhat} = j \ket{\nhat}$. In terms of stereographic
coordinates
\beq
z = \tan{\tshf}\tta e^{i\phi}, \quad
\zbar = \tan{\tshf}\tta e^{-i\phi}, 
\eeq
the state may also be written as $\ket z = \exp(z\Jmi) \ket{\zhat}$.
Note that $\ket z$ is not normalized, and
$\tran{z}{z} = (1+\zbar z)^{2j}$. We shall write $\ket{\nhat}$, $\ket z$,
or $\ket{\tta,\phi}$ interchangeably as needed. Secondly, we shall need to
discuss points on the complex unit sphere, for which $\tta$ and $\phi$ are
not real, or equivalently, $\baz$ and $z$ are not true complex conjugates.
Such points are specified by the pair $(\baz,z)$.

Our starting point is the
semiclassical approximation \cite{spg+etal} to the propagator
$K = \mel{z_f}{e^{-i\ham T}}{z_i}$:
\beq
\Ksc(\bzf,z_i; T)
   = \sqrt{{N\by 2j}} \lp {\ptl^2S \by \ptl\bzf \ptl z_i} \rp^{1/2}
       \exp\lp S + {i\by 2}\int_0^T Adt \rp.
                   \label{semiprop}
\eeq
with,
\bea
S &=& j\ln N
           + \int_0^T \lsb j{\dot\zbar z - \zbar \dot z \by 1 +\zbar z}
                        -i \Hsc(\baz,z) \rsb dt, \\
A &=& {\ptl \by \ptl\baz} {(1+\baz z)^2\by 4j}
                 {\ptl\Hsc \by \ptl z} + z \tofro \baz, \\
N &=& (1+\bzf z(T))(1 + \baz(0) z_i).
\eea 
Here, $S$ is  the action along the least action trajectory
from the point $(\baz(0),z_i)$ to $(\bzf, z(T))$; as noted in Ref.
\onlinecite{spg+etal}, to find such a trajectory, i.e., a solution of the
Euler-Lagrange equations,
\beq
\dot\zbar = {i\by 2j} (1+\zbar z)^2 {\ptl\Hsc \by \ptl z}, \quad
\dot z = -{i\by 2j} (1+\zbar z)^2 {\ptl\Hsc \by \ptl\zbar},
    \label{ELeqns}
\eeq
we must allow $\zbar(0) \ne \bzi$
and $\bzf \ne \overline{z(T)}$. The integral in $S$ must be evaluated
along this trajectory. So must $\int A dt$, the Solari-Kochetov correction,
which is $O(1/j)$ relative to $S$, and key to getting a propagator that
is self-consistently replicating under composition. Further,
\beq
\Hsc(\zbar, z) = \mel{z}{\ham}{z},
\eeq
which we shall call the semiclassical Hamiltonian. Finally, we must sum
the the right hand side in \eno{semiprop} over all solutions of the equations
of motion if there is more than one.

Our goal is to find the Green function, and then the energy spectrum by
looking for its poles. In step 1, we evaluate the Laplace transform
\beq
\Fsc(\zbf, z_i; E) = \int_0^T \Ksc(\zbf, z_i; T) e^{iET} dT
\eeq
by the stationary phase method. This naturally leads to the action at
fixed energy $W(E) = S(T) + iET$,
where $E$ and $T$ are related by 
\beq
T(E) = -i {\ptl W(E) \by \ptl E}, \quad E(T) = i{\ptl S(T) \by \ptl T}.
    \label{EvsT}
\eeq
Then, using known methods \cite{marinov80}, we can show that
\beq
\Fsc = \left. e^{i\gam} \sqrt{{N\by 2j}}
          \lp {2\pi \by {\dot z}_i{\dot\zbar}_f} \rp^{1/2}
            \exp\lp W + {i\by 2}\int A dt \rp \right|_{T = T(E)}.
\eeq
Here we have included a Maslov-like phase $e^{i\gam}$, which arises
because the mapping from $T$ to $E$ is many to one.
%
%
%
%
%
%
%
This is best seen by considering the case $z_f = z_i$, which we shall
shortly encounter when we take the trace. Now, for the same
energy $E$, there is more than one trajectory corresponding to multiple
traverses of the same fundamental orbit. So, $T$ is an integer
multiple of the basic period $T_0$, the different branches of $W$
differ by additive integer multiples of $iET_0$, and the phase $e^{i\gam}$
is $(-1)^n$ for $n$ traverses.

In step 2, we again use stationary phase to perform the trace that gives
the semiclassical Green function:
\beq
\Gsc(E) = {2j + 1 \by \pi}\int {dz d\zbar \by (1+\zbar z)^{2j+2}}
                              \Fsc(\zbar, z; E).
\eeq
The stationary phase condition
yields $z(T) = z$, and $\zbar(0) = \zbar$, so now we are only considering
closed classical orbits with energy $E$, with momenta that match smoothly
at the end points. Next, we must integrate over small fluctuations $\eta$
and $\bar\eta$ in $z$ and $\bar z$. For the fundamental orbit at energy $E$,
we can write
\beq
\Gsc = -\sqrt{\jtil \by \pi}
           \int {d\eta d\bar\eta \by (1+\zbar z)} 
                 {1\by |\dot z|} \exp(\Zsc) \exp(-{\tshf}Q)
\eeq
using known classical mechanical identities. Here,
\bea
\Zsc &=&  \int_0^T \lsb \jtil {\dot\zbar z - \zbar \dot z \by 1 +\zbar z} 
           + i {(1+\baz z)^2\by 4j}
                {\ptl^2\Hsc \by \ptl\baz \ptl z} \rsb dt, \label{defZ}\\
Q &=& {2j +1 \by (1+\zbar z)^2}
        \lp 2\bar\eta\eta -{\dot z \by \dot\zbar}{\bar\eta}^2
                          -{\dot\zbar \by \dot z}\eta^2 \rp.
\eea
The quadratic form $Q$ has a zero eigenvalue, corresponding to a
fluctuation $(\bar\eta,\eta)$ that moves us along the orbit. If we parametrize
this direction by the time $t$, and the orthogonal direction by a variable
$r$, then,
\beq
\lp\begin{array}{c} \eta \\ \bar\eta \end{array}\rp
=  \lp\begin{array}{c} \dot z \\ \dot\zbar \end{array}\rp t
    + \lp\begin{array}{c} \dot z \\ -\dot\zbar \end{array}\rp r,
\eeq
and $Q = 4r^2 |\dot z|^2$. The integration over all points along the
orbit turns into an integral over $t$ from $0$ to $T$, and with the
jacobian $\ptl(\eta,\bar\eta)/\ptl(t,r) = -2|\dot z|^2$, the $r$ integral
may also be done. The result is that for one traverse of the orbit we get 
$-iT \exp(\Zsc)$. For $n$ traverses, we multiply by $(-1)^n$ and let
$\Zsc \to n \Zsc$. Summing over all $n$, we get
\beq
\Gsc(E) = {iT(E) \by 1 + \exp(-\Zsc(E))}.
\eeq
This has poles (each with unit residue) whenever
\beq
\Zsc(E) = (2n+1)i\pi. \label{bsrule2}
\eeq

\Eno{bsrule2} is our quantization condition. The first term in \eno{defZ}
is $-i\jtil$ times the area $\Asc$. Unlike $\Ac$, $\Asc$ is computed on
the orbit $\Hsc = E$, not $H_c = E$, since it is $\Hsc$ that appears in
\eno{ELeqns}. In the second term, we note that
\beq
(1+\zbar z)^2 \mixpar{\Hsc}{z}{\zbar} = \grad^2_{\Om}\Hsc, \label{LapH}
\eeq
where $\grad^2_{\Om}$ is the angular part of the Laplacian.

A true BS rule must be in terms of the classical Hamiltonian, $H_c$, not
$\Hsc$. To cast \eno{bsrule2} into the form (\ref{bsrule}), we must relate
$H_c$ and $\Hsc$. Using \eno{ylmgenfn}, we can show that
\beq
\mel{\nhat}{\ylmop(\bJop)}{\nhat}
   = \lsb 1 - {\ell(\ell -1) \by 4j} + O(j^{-2})\rsb \ylm(\bJ),
\eeq
with $\bJ = j\nhat$ \cite{exactcoeff}. Since,
$\grad^2_{\Om}\ylm(\bJ) = -\ell(\ell+1)\ylm(\bJ)$, and
$\bJ\cdot\grad_J\ylm(\bJ) = \ell\ylm(\bJ)$, it follows that
$\Hsc = H_c + H_1$, where
\beq
H_1 = {1\by 4j}\grad^2_{\Om}H_c + {1\by 2j} \bJ\cdot\grad_J H_c.
     \label{defH1}
\eeq
We have written $H_c$ instead of $\Hsc$ on the right since the corrections
are of $O(j^{-2})$.

The difference between $\Asc$ and $\Ac$ is now easy to find. Let us denote
the solutions of \eno{ELeqns} $(\bzsc,\zsc)$ and those with $H_c$ instead
of $\Hsc$ by $(\zbar_c,z_c)$. If we regard $\bzsc$ and $z_c$ as functions
of $E$ and $z$, then
\bea
H_c\bigl(\bzsc(E,z), z\bigr) &=& E - H_1, \\
H_c\bigl(\zbar_c(E,z), z\bigr) &=& E.
\eea
Writing $\bzsc = \zbar_c + \zbar_1$, $\zsc = z_c + z_1$, and using the
equations of motion, we find
\beq
\zbar_1 = i{(1+\zbar z)^2 \by 2j} {H_1 \by \dot z}, \quad
    z_1 = -i{(1+\zbar z)^2 \by 2j} {H_1 \by \dot\zbar}.
\eeq
In this way we find
\beq
\jtil \Asc = \jtil \Ac + \int_0^T H_1 dt.
\eeq
Substituting this in \eno{defZ}, and making use of \etwo{LapH}{defH1}, we
see that \etwo{bsrule2}{bsrule} are equivalent.

We mention in closing that we can also use coherent states to find a BS rule
for one dimensional potential problems. The result is akin to \eno{bsrule2}.
To show its
equivalence to the textbook rule, we use Weyl ordering and the Wigner-Moyal
formalism to relate $\ham$, $H_c$, and $\Hsc$. In the spin case,
the problems of ordering, and of relating $H_c$ to $\Hsc$, are neatly
solved by the generating formula (\ref{ylmgenfn}).

This work is supported by the National Science Foundation Grant Nos.
DMR-0202165 (AG) and DMR-0132990 (MS).

%
%

\begin{thebibliography}{99}
\bibitem{qtmcorrect} We regard a BS rule as proper only if it includes
these corrections. For a single particle, e.g., the action
for an orbit must be equated to $(n+\hf)(2\pi\hbar)$, and the rule would
be improper if the $\hf$ were left out. The necessity and rationale
for including these corrections is well known. In the spin case, we
similarly require the $1/j$ corrections relative to leading order.
\bibitem{degreepoly} The degree of this polynomial can be no greater than
$2j$, but in physical situations it is usually much smaller.
\bibitem{areasign} Since every closed curve divides the sphere into
two parts, we fix the orientation (indicated by the suffix +) on the sphere
such that ${\cal A}_c^+$ is the area {\it inside\/} the orbit traversed in
the counterclockwise sense.
\bibitem{exactcoeff} The exact coefficient on the right 
is $(\ell!/2^{\ell}j^{\ell}){2j \choose \ell}$.
\end{thebibliography}
\end{document}